


\def\NI{\noindent}
\magnification=\magstep 1
\overfullrule=0pt
\hfuzz=16pt
\voffset=0.0 true in
\vsize=8.8 true in
   \def\NP{\vfil\eject}
   \baselineskip 20pt
   \parskip 6pt
   \hoffset=0.1 true in
   \hsize=6.3 true in
\nopagenumbers
\pageno=1
\footline={\hfil -- {\folio} -- \hfil}
\headline={\ifnum\pageno=1 \hfill October 1992 \fi}

\hphantom{AA}

\hphantom{AA}

\centerline{\bf EXACT SOLUTION OF A PHASE SEPARATION MODEL}
\centerline{\bf WITH CONSERVED ORDER PARAMETER DYNAMICS}

\vskip 0.4in

\centerline{\bf Vladimir~Privman}

\vskip 0.2in

{\NI}{\sl{Department of Physics, Clarkson University, Potsdam, New
York \ 13699--5820, USA}}

\vskip 0.4in

\NI {\bf PACS:}$\;$ 82.20.-w, 05.40.+j

\vskip 0.4in

\centerline{\bf ABSTRACT}

Pairwise particle-exchange model on a linear lattice is solved
exactly by a new rate-equation method. Lattice sites are occupied by
particles A and B which can exchange irreversibly provided the local
energy in reduced. Thus, the model corresponds to a
zero-temperature Kawasaki-type phase separation process. Due to
local order-parameter conservation, the dynamics reaches a frozen
state at large times, the structure of which depends on the initial
conditions.

\NP

Recently there has been much interest in modeling phase separation
and spinodal decomposition [1] by simple irreversible, effectively
zero-temperature low-dimensional  stochastic dynamical systems
[2-4]. Specifically, some variants of nonconserved order parameter
dynamical models in $D=1$, corresponding effectively to $T=0$
Glauber-type spin systems, have been solved exactly for properties
such as the structure factor and average domain size (as functions
of time, $t$); see [4] for details. The underlying mechanism leading
to cluster growth in $D=1$ is pairwise annihilation of interfaces
separating ordered domains. The interfacial motion is diffusional
and it corresponds also to certain diffusion-limited particle
annihilation models [4-12].

The $T \to 0$ limiting model involves interface annihilation which
is a process lowering the local energy and therefore has Boltzmann
factor $+\infty$ associated with its transition probability at $T =
0$. Interface diffusion does not change the local energy and
therefore has Boltzmann factor $1$. Finally, interface generation
(birth) has Boltzmann factor $0$ (due to energy cost) at $T=0$. The
$T=0$ models referred to earlier, correspond to allowing for both
annihilation and diffusion. However, one could also consider
processes with interface annihilation only. There has been limited
discussion of models of such stationary annihilating particles
(interfaces in the phase separation nomenclature) in the literature
[13-15]. Specifically, exact $D=1$ results can be obtained [13].

The phase separation process in the latter,
diffusionless case does not continue indefinitely (as it does in
the annihilation-with-diffusion models). Indeed, considering $D=1$
for instance, one can easily visualize that when interfaces
initially in ``contact'' are depleted by annihilation, the resulting
configuration still contains some isolated ``unreacted''
interfaces. Thus the system will actually freeze in a certain
partially ordered state which depends on the initial state, --- a
direct manifestation of the irreversible (nonergodic) nature of the
$T=0$ dynamics.

Let us now turn to conserved order parameter, spin- or
particle-exchange Kawasaki-type dynamical models. There are several
differences as compared to the nonconserved models just surveyed.
Notably, interfacial processes even at $T=0$ are more complicated
for the conserved case. Specifically, let us consider the $D=1$
binary AB-mixture model: each site of the $1D$ lattice is occupied
by particle A or particle B. The locally conserved order parameter
is the difference of the A-\ and B-particle densities.

Nearest-neighbor particle exchanges can lower local energy (reduce
number of interfaces) in the following configurations:

$$ \ldots {\rm ABAB} \ldots \to \ldots {\rm AABB} \ldots \; ,
\eqno(1) $$

$$ \ldots {\rm BABA} \ldots \to \ldots {\rm BBAA} \ldots \; .
\eqno(2) $$

\NI Note that three interfaces (A--B or B--A bonds) ``reacted'' to
yield one interface. Particle exchanges that do not change energy
locally are possible in configurations like

$$ \ldots {\rm AABA} \ldots \to \ldots {\rm ABAA} \ldots \; ,
\eqno(3) $$

\NI and three similar reflected and/or relabeled
(A$\leftrightarrow$B) configurations. Here hopping of an interface
must be mediated by the presence of another, nearby interface.

Thus interfacial dynamics in the particle-exchange models is more
involved than in the nonconserved case, even at $T=0$. Specifically,
energy-conserving interfacial motion, (3), is no longer simple free
diffusion. Thus, freezing rather than full phase separation occurs
asymptotically for large times in models with both energy-lowering
and energy-conserving moves allowed, or with only energy-lowering
moves allowed. In the former case the $D=1$ frozen state contains
single interfaces while in the latter case both single interfaces
and pairs of interfaces are ``frozen in.'' Several numerical studies
were reported [16-19] of such particle-exchange models for $D$ up to
5. As in the nonconserved case, some of the properties of the $D=1$
models are different from $D>1$. However, the general expectation of
the ``freezing'' of the domain structure at large times applies, for
conserved dynamics, at all $D$.

Derivation of exact $D=1$ results for conserved-dynamics models
proved considerably more difficult than for the nonconserved models.
Palmer and Frisch [18] developed a method by which the asymptotic
($t=\infty$) density of the residual, frozen in interfaces for the
dynamics with energy-lowering moves only, (1)-(2), can be
calculated, starting from the initially fully ``mixed''
alternating-AB state. More recently, Elskens and Frisch [20]
extended this approach to allow calculation of the rate of approach
to the freezing density, and also obtain results for the initial
state randomly populated by the equal-probability mixture of A and B.

In the present work, we present the full time-dependent solution of
the $D=1$ model. We use a new method inspired by techniques
developed in studies of $D=1$ random sequential adsorption [21-23].
The solution is obtained for both alternating and random initial
conditions, and it recovers all the asymptotic results of [18,20].

We consider a linear lattice each site of which is occupied by
either particle A or particle B. Initially, the particle arrangement
is alternating (denoted $alt$), $\ldots$ABABABA$\ldots$, or random
(denoted $ran$). In the former case we neglect end effects;
all our calculations will be for the infinite linear
lattice (as opposed to [18,20]). In the random initial case, we
assume that each site is either A or B with equal probability, at
time $t=0$, so that the initial densities of A and B species are
50\%.

The dynamics consists of particle exchanges which decrease the
number of ``broken'' bonds (interfaces) A--B and B--A. Thus the
allowed moves are (1) and (2), i.e., only pairs of particles
centered in a fully alternatively-ordered group of 4, can exchange
thus reducing the local number of interfaces from 3 to 1. We assume
asynchronous, continuous-time dynamics: each allowed-configuration
nearest-neighbor pair AB or BA undergoes the exchange process, i.e.,
the particles switch their lattice sites, at the rate $R$,
independent of other exchange events. As is well known, in the
continuum-time limit of asynchronous dynamics one can disregard
interference of exchanges of pairs which share a site, such as the
two pairs sharing the central A site in $\ldots$ABABA$\ldots\;$.
The rate parameter $R$ will be incorporated in the time variable so
that we denote the physical product $Rt$ simply by $t$ (effectively
setting $R=1$).

Our aim is to calculate the density of interfaces, $I(t)$, i.e., the
fraction of A--B and B--A bonds, as a function of time. Initially we
have

$$ I_{alt}(0)=1 \qquad {\rm and} \qquad I_{ran}(0)={1 \over 2} \; .
\eqno(4) $$

Our method of solution involves calculation of probabilities $P
(k , t)$ that a randomly selected continuous group of $k \geq 3$
lattice sites is fully alternatively-ordered, i.e., that it is
occupied, at time $t$, by $k$ particles $ABAB\ldots$ or
$BABA\ldots$, where the sequence is alternating and $k$-site long.
Note, however, that we do not impose any condition on the
configuration outside this group. Thus, the alternatively ordered
region need not be exactly $k$-site long, and in fact it may be
part of a longer alternating sequence of sites, at one or both
ends. In this respect our definition differs from earlier works and
follows instead the ideas developed for random sequential
adsorption models [21-23].

The rate of decrease of the interface density is given by

$$ - {dI (t)  \over dt} = 2 P (4,t) \; , \eqno(5) $$

\NI where the factor 2 accounts for the reduction by 2 (from 3 to 1)
of the local interface number in each exchange event. The factor
$P(4,t)$ is the probability that a randomly selected group of 4
sites is one of the allowed-exchange configurations, (1) or (2). As
already mentioned, at $t=\infty$ only single isolated interfaces and
isolated pairs of nearest-neighbor interfaces survive in a frozen
state. The number of interfaces which are paired up is in fact given
simply by

$$ I^{(paired)} (\infty) = 2 P(3, \infty ) \; , \eqno(6) $$

\NI while the number of single interfaces in the final state can be
calculated as the difference $I(\infty)-I^{(paired)}(\infty)$.
Indeed, only the probability $P(3,t)$ remains finite as $t \to
\infty$; all $P(k>3,t)$ vanish in the large-$t$ limit; see explicit
results below.

The decrease of the probability $P(k,t)$ with time is governed by
the following rate equation,

$$ - {d P(k,t) \over dt } = (k-3) P(k,t) +2 P(k+1,t) + 2 P(k+2,t) \;
, \qquad (k \geq 3) \; . \eqno(7) $$

\NI The first term is the rate at which the alternating order is
disrupted by pairwise exchanges of pairs such that their
``defining'' 4 sites, i.e., the original pair sites and the two
neighbor sites on both sides, fall fully within the $k$-group under
consideration. The second term corresponds to exchange events of the
pairs which are located at the two ends of the $k$-group. Indeed,
for an end pair to lie within 4-sequences (1) or (2) which
correspond to the allowed-exchange configurations, our $k$-group must
in fact be part of a larger ordered group, of length $(k+1)$,
including all the four ``deciding'' sites of a given end-pair. (One
of these sites is external to the original $k$-group.) Finally, the
third term corresponds to disruption of alternating order due to
exchange of the end sites of the $k$-group with their nearest
neighbors just outside the $k$-group. For one of these two pairs of
sites which are half-external to the $k$-group to exchange, the four
``deciding'' sites require consideration of a $(k+2)$-site-long
extended group.

The rate equations (7) form a relatively simple hierarchy only in
$D=1$. For $D>1$, $k$-groups are replaced by more complicated
clusters and no exact solution is possible by this method.
The exact rate equations (7) must be solved with the following
initial conditions,

$$ P_{alt}(k,0) = 1 \qquad {\rm and} \qquad P_{ran}(k,0)=2^{1-k} \;
. \eqno(8) $$

The solution can in principle be obtained by the generation function
method. However, a much simpler way is to notice that the hierarchy
(7) is solved by the Ansatz

$$ P(k,t)=P(k,0) Q(t) e^{-(k-3)t} \; , \eqno(9) $$

\NI for both alternating and random initial conditions (but not
generally). A straightforward calculation then yields

$$ Q_{alt}(t) = \exp \left( 2 e^{-t} +  e^{-2t} -3 \right)
\qquad {\rm and} \qquad
Q_{ran}(t) = \exp \left(  e^{-t} + {1 \over 4}{ e}^{-2t} -{5 \over
4} \right) \; . \eqno(10) $$

Collecting all the results and definitions, (4)-(10), and solving for
$I(t)$, yields, after some algebra, the results

$$ I_{alt} (t) = 1 -{2 \over { e}^4} \int\limits^2_{1+{ e}^{-t}} {
e}^{z^2} dz \; , \eqno(11) $$

$$ I_{alt} (\infty) = 1 -{2 \over { e}^4} \int\limits^2_1 { e}^{z^2}
dz \simeq 0.450898 \; , \eqno(12) $$

$$ I_{alt}^{(paired)} (\infty) = {2 \over { e}^3} \simeq 0.099574 \;
, \eqno(13) $$

$$ I_{ran} (t) = {1 \over 2} - {1 \over 2 e^{9/4} }
\int\limits^{3/2}_{1+{1 \over 2} { e}^{-t}} { e}^{z^2} dz \; ,
\eqno(14) $$

$$ I_{ran} (\infty) = {1 \over 2} -{1 \over 2 e^{9/4} }
\int\limits^{3/2}_1 { e}^{z^2} dz \simeq 0.362957 \; , \eqno(15) $$

$$ I_{ran}^{(paired)}  (\infty) = {1 \over 2{ e}^{5/4}} \simeq
0.143252 \; . \eqno(16) $$

The functions (11) and (14) are plotted in Figure~1. To facilitate
comparison, their values were normalized by $I(0)$, see (4). The
most profound feature of the irreversible dynamics leading to frozen
states is of course the dependence of the final state on the initial
conditions. Explicit results (11)-(16) are, to author's knowledge,
the first exact time dependent expressions available for
conserved-dynamics models. It is hoped that the method of solving
the $D=1$ models will be used to obtain additional exact
$1D$ results as well as new approximation schemes in $D>1$.

The author wishes to thank Professor H.L.~Frisch for helpful comments and
suggestions.

\NP

\centerline{\bf REFERENCES}

{\frenchspacing

\item{1.} Review: J.D. Gunton, M. San Miguel
and P.S. Sahni, in {\sl Phase
Transitions and Critical Phenomena}, Vol. 8, C. Domb and J.L.
Lebowitz, eds. (Academic Press, New York, 1983), p. 269.

\item{2.} M. Scheucher and H. Spohn, J. Stat. Phys. {\bf 53}, 279
(1988).

\item{3.} B. Hede and V. Privman, J. Stat. Phys. {\bf 65}, 379
(1991).

\item{4.} V. Privman, J. Stat. Phys. (1992), in print.

\item{5.} M. Bramson and D. Griffeath, Ann. Prob. {\bf 8}, 183
(1980).

\item{6.} D.C. Torney and H.M. McConnell, J. Phys. Chem. {\bf 87},
1941 (1983).

\item{7.} Z. Racz, Phys. Rev. Lett. {\bf 55}, 1707 (1985).

\item{8.} A.A. Lushnikov, Phys. Lett. {\bf 120}A, 135 (1987).

\item{9.} M. Bramson and J.L. Lebowitz, Phys. Rev. Lett. {\bf 61},
2397 (1988).

\item{10.} D.J. Balding and N.J.B. Green, Phys. Rev. A{\bf 40}, 4585
(1989).

\item{11.} J.G. Amar and F. Family, Phys. Rev. A{\bf 41}, 3258
(1990).

\item{12.} A.J. Bray, J. Phys. A{\bf 23}, L67 (1990).

\item{13.} V.M. Kenkre and H.M. Van Horn, Phys. Rev. A{\bf 23}, 3200
(1981).

\item{14.} H. Schn\"orer, V. Kuzovkov and A. Blumen, Phys. Rev.
Lett. {\bf 63}, 805 (1989).

\item{15.} H. Schn\"orer, V. Kuzovkov and A. Blumen, J. Chem. Phys.
{\bf 92}, 2310 (1990).

\item{16.} A. Levy, S. Reich and P. Meakin, Phys. Lett. {\bf 87}A,
248 (1982).

\item{17.} P. Meakin and S. Reich, Phys. Lett. {\bf 92}A, 247 (1982).

\item{18.} R.G. Palmer and H.L. Frisch, J. Stat. Phys. {\bf 38}, 867
(1985).

\item{19.} F.C. Alcaraz, J.R. Drugowich de Fel\'icio and R.
K\"oberle, Phys. Lett. {\bf 118}A, 200 (1986).

\item{20.} Y. Elskens and H.L. Frisch, J. Stat. Phys. {\bf 48}, 1243
(1987).

\item{21.} E.R. Cohen and H. Reiss, J. Chem. Phys. {\bf 38}, 680
(1963).

\item{22.} J.J. Gonz\'alez, P.C. Hemmer and J.S. H{\o}ye, Chem.
Phys. {\bf 3}, 228 (1974).

\item{23.} Review: M.C. Bartelt  and V. Privman, Int. J. Mod. Phys.
B{\bf 5}, 2883 (1991).

}

\vfill

\NI {\bf Figure~1.\ \ }Lower curve: density of interfaces, $I(t)$,
for alternating initial conditions; see (11). Upper curve: density of
interfaces for random initial conditions; see (14). The curves
plotted are $I(t)/I(0)$, where the initial densities are given
in (4).

\eject\end